\documentclass[twocolumn,english,aps,prb,superscriptaddress,floatfix]{revtex4}
\usepackage{color}
\usepackage{amsmath}
\usepackage{graphicx}
\usepackage{amssymb}
\usepackage{hyperref}
\usepackage{dsfont}
\allowdisplaybreaks
\makeatletter
\@ifundefined{textcolor}{}
{%
 \definecolor{BLACK}{gray}{0}\definecolor{WHITE}{gray}{1}
 \definecolor{RED}{rgb}{1,0,0}
 \definecolor{GREEN}{rgb}{0,1,0}
 \definecolor{BLUE}{rgb}{0,0,1}
 \definecolor{CYAN}{cmyk}{1,0,0,0}
 \definecolor{MAGENTA}{cmyk}{0,1,0,0}
 \definecolor{YELLOW}{cmyk}{0,0,1,0}
 }

\@ifundefined{definecolor}
 {\usepackage{color}}{}
\@ifundefined{definecolor}{\@ifundefined{definecolor}
 {\usepackage{color}}{}
}{}

\newcommand{\ba}{\begin{eqnarray*}}
\newcommand{\ea}{\end{eqnarray*}}
\newcommand{\baa}{\begin{eqnarray}}
\newcommand{\eaa}{\end{eqnarray}}
\newcommand{\bea}{\begin{eqnarray}}
\newcommand{\eea}{\end{eqnarray}}
\newcommand{\be}{\begin{equation}}
\newcommand{\ee}{\end{equation}}
\newcommand{\dis}{\displaystyle}
\newcommand{\fract}[2]{\frac{\dis #1}{\dis #2}}
\newcommand{\eqn}[1]{(\ref{#1})}
\newcommand{\ket}[1]{\mid\! #1\rangle}
\newcommand{\bra}[1]{\langle #1\!\mid} 
\newcommand{\ep}{{\epsilon}}
\newcommand{\hH}{{\hat{H}}}
\newcommand{\mE}{{\mathcal{E}}}
\newcommand{\dagga}{{\phantom{\dagger}}}

\makeatother

\usepackage{babel}

\begin{document}

\title{
Nanomechanical dissipation at a tip-induced Kondo onset 
}

\author{Pier Paolo Baruselli}
\affiliation{International School for
  Advanced Studies (SISSA), Via Bonomea
  265, I-34136 Trieste, Italy}
\author{Michele Fabrizio}
\affiliation{International School for
  Advanced Studies (SISSA), Via Bonomea
  265, I-34136 Trieste, Italy}
\author{ Erio Tosatti}
\affiliation{International School for
  Advanced Studies (SISSA), Via Bonomea
  265, I-34136 Trieste, Italy}
\affiliation{CNR-IOM Democritos, Via Bonomea 265, 34136 Trieste, Italy}
\affiliation{International Centre for Theoretical Physics (ICTP), Strada Costiera 11, I-34151 Trieste, Italy}



\date{\today}
\begin{abstract}
The onset or demise of Kondo effect in a magnetic impurity on a metal surface can be triggered, as sometimes
observed, by the simple mechanical nudging of a tip. Such a 
mechanically-driven quantum phase transition must reflect in a corresponding 
mechanical dissipation peak; yet, this kind of
signature
has not been focused upon so far. 
Aiming at the simplest theoretical modelling, we treat the impurity as an Anderson impurity model, 
the tip action as a hybridization switching, 
and solve the problem by numerical renormalization group. Studying this model as function of temperature and magnetic field we are able to isolate the Kondo contribution to dissipation. 
While that is,  reasonably, of the order of the Kondo energy, its temperature evolution shows a surprisingly large tail even above the Kondo temperature. 
The detectability of Kondo mechanical dissipation in atomic force microscopy is also discussed.  
\end{abstract}

\maketitle

\section{Introduction}

\begin{figure}[hbt]
\includegraphics[width=0.45\textwidth]{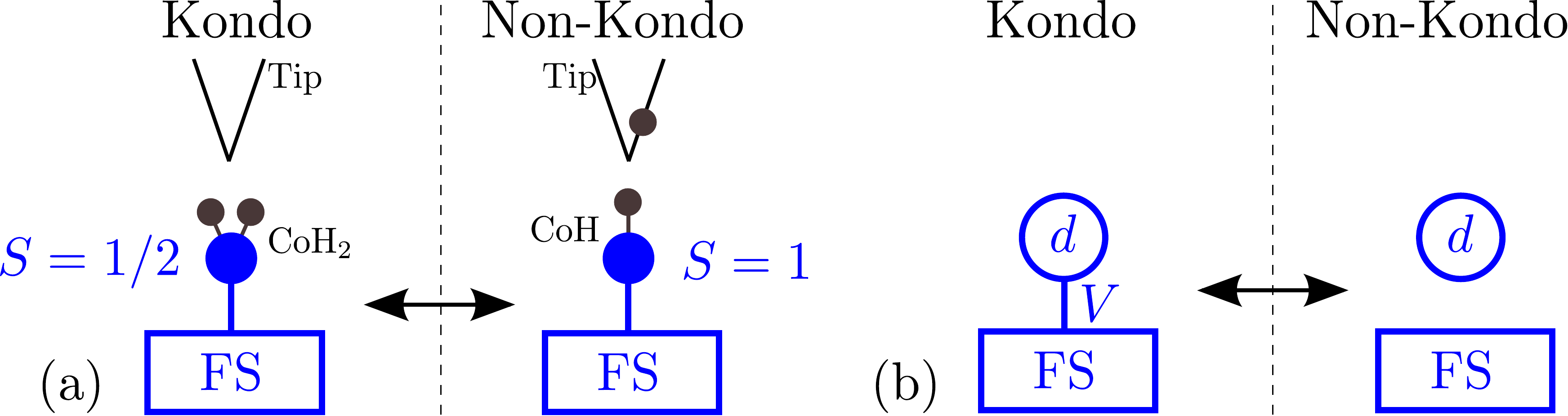}
\caption{(a) In experiments such as Refs. \onlinecite{Jacobson2015,Jacobson2016}, the Kondo effect is reversibly switched on and off by binding and unbinding an H atom to a CoH impurity complex.
(b) Model of the present calculation, where we switch the hopping $V$ between an impurity orbital $d$ endowed with an electron with S=1/2, and a Fermi sea (FS), therefore turning on and off the Kondo screening. 
This cartoon does not exactly represent the same physics as in (a), but
we will show that it does permit the calculation of the temperature and magnetic field dependence of energy dissipation caused by switching the Kondo screening on and off. 
}\label{fig_tip}
\end{figure}

The dissipation characteristics
of nanomechanical systems such as an oscillating AFM or STM tip above a surface is increasingly
emerging as a local spectroscopic tool. The exquisite sensitivity of these systems permits the study 
of physical phenomena, in particular of phase transitions, down to the nanoscale. ~\cite{Ternes2009, Kisiel2011, Langer2014, Kisiel2015, Jacobson2016} 
Tip-induced internal transitions in quantum dots have been known to cause sharp dissipation peaks. ~\cite{Cockins_PNAS_2010} 
The injection by pendulum AFM of 2$\pi$ slips in the surface phase of a charge-density-wave,  an event akin to a local first order 
classical phase transition, is also marked by dissipation peaks.~\cite{Langer2014} 
Parallel to, but independent of that, many-body electronic and magnetic effects at the nanoscale were in the last decades  studied with atomic resolution by STM. 
The Kondo effect \cite{Kondo1964,Hewson} -- the many-body correlated state of a magnetic impurity --  is commonly probed on metallic surfaces by STM tips, 
where it gives rise to a prominent zero-bias anomaly. \cite{schneider_ceag,madhavan_coau,Ternes2009, Jacobson2016}. 
Thus far, these two phenomena, nanomechanical dissipation peaks on one hand, and the onset or demise of Kondo screening on the other hand -- itself a many-body quantum phase transition -- have not been connected.  
Yet, experimentally the onset and offset of a Kondo state, has been mechanically triggered in several instances. 
Recent data\cite{Zhao1542,Bogani2008,choi_kondoswitch,Komeda2011,Wagner2013,Jacobson2015,Jacobson2016} 
show that it is possible to selectively switch on and off the Kondo effect via atomic manipulation, 
a switching detected through the appearance and disappearance of a zero bias STM conductance anomaly. 
As a typical example, a surface  impurity may be switched by a tip action from a spin $1/2$ state, which is Kondo screened,  to a $S=1$ or a $S=0$ state where the Kondo effect  disappears- see Fig. \ref{fig_tip},
or viceversa. In more general circumstances the Kondo temperature $T_K$ of the impurity can be mechanically manipulated\cite{Liang2002,Neel2007,iancu_tkchange}.
Also, Kondo underscreening can be achieved  via mechanical control of a break junction \cite{Parks11062010}. 
The conductance anomalies across Kondo impurities have a rich literature, both experimental and theoretical,
reviewed in part in Refs. \onlinecite {natelson_review, requist_review}. 

Here we provide the first theoretical prediction of the much less studied nanomechanical dissipation at  the onset and demise of Kondo effect.
We will do that in the simplest, prototypical case inspired by, e.g., a  slowly vibrating tip near a surface-deposited magnetic impurity or molecular radical.
The tip mechanical Kondo dissipation magnitude per cycle and its characteristic temperature dependence 
have not yet been measured so far.
Similar and complementary to conductance anomalies, we expect 
the dissipation to involve Kondo energy scales, and to be influenced and eventually disappear upon increasing temperature and/or magnetic field. 

The possibility of dissipation and/or friction associated to the Kondo effect is actually not new. Earlier studies addressed different conditions, 
typically  atoms moving near metal surfaces \cite{PhysRevB.60.5969,PhysRevB.58.2191},  as well as  quantum dots \cite{PhysRevB.61.2146,goker_thermopower_kondo}. Kondo dissipation was also
addressed in the linear response regime via the dynamic charge susceptibility \cite{PhysRevB.55.2578}, in connection with ac-conductance \cite{PhysRevB.61.2146}, and thermopower \cite{goker_thermopower_kondo}.
Our work agrees with the previous general result that the Kondo effect enhances the total dissipation. The present method is non-perturbative and numerically exact, while its validity is restricted to zero frequency, adequate to describe dissipation of a tip, whose mechanical motion is slow.  In addition, our approach naturally allows for the inclusion of magnetic fields and of further degrees of freedom. We will come back the non-equilibrium finite-frequency problem in the future.

After a presentation of the energy dissipation modeling in Section \ref{sec_model}, 
we show that the dissipation caused by a sufficiently slow switching on and off of the hybridization between an impurity and a Fermi sea can be reduced to a ground state calculation. 
The many-body Kondo physics is then introduced by solving as a function of temperature
an Anderson model by numerical renormalization group (NRG). 
In this way we arrive to a dissipation per cycle whose peak value as temperature increases is initially proportional to the hybridisation magnitude $\Gamma$, 
then decreasing from  $T=T_K$ upwards roughly as $-T_K (2/\pi) |\log[1 +4 (T/T_K)^2]|$,
until reaching the ultra-high temperature $T \gg  \Gamma$, where dissipation drops as  $-\Gamma |\log{ T}|$ (Section \ref{sec_nrg}).  
The Kondo-related dissipation drops, between $T=0$ and $T=T_K$ by about $\Delta E_\text{diss} \sim T_K$ per cycle, 
but interestingly a very important dissipation extends well above  $T_K$.
Additionally revealing is the effect of a large external magnetic field, which produces a drop of the low temperature, $T\ll T_K$, energy dissipation again proportional to  $T_K$, 
yet through a  factor  $\Delta E_\text{diss}/ T_K$ much larger than 1 due to 
$\log T_K$ corrections. 

As an independent check of these Anderson model results, where $\Gamma$ is the only quantity that can be switched, we then repeat calculations of dissipation in a Kondo model, 
where the exchange coupling $J$ can be switched on and off, mimicking the onset and demise of Kondo screening. The results support those of the Anderson model,
and permit in addition to study the dissipation of a ferromagnetic Kondo impurity.
The observability of Kondo dissipation  against the background mechanical dissipation of different origin is discussed at the end(Section \ref{sec_conclusions}).

\section{Model}\label{sec_model}
We will not attempt to describe in detail the mechanically-driven quantum phase transition, and model instead its effect as a quantum quench by assuming
a model Hamiltonian $\hH$ that can be switched from $\hH_0$ (no Kondo) to $\hH_1$ (Kondo-like) and viceversa. 
The change of dissipation calculated in this manner, in particular as a function of temperature and magnetic field, will clearly identify the Kondo-related dissipation, such as that would be mechanically measured by the damping of a tip whose action was to cause the onset or demise of Kondo. 

\begin{figure}[bt]
\includegraphics[width=0.45\textwidth]{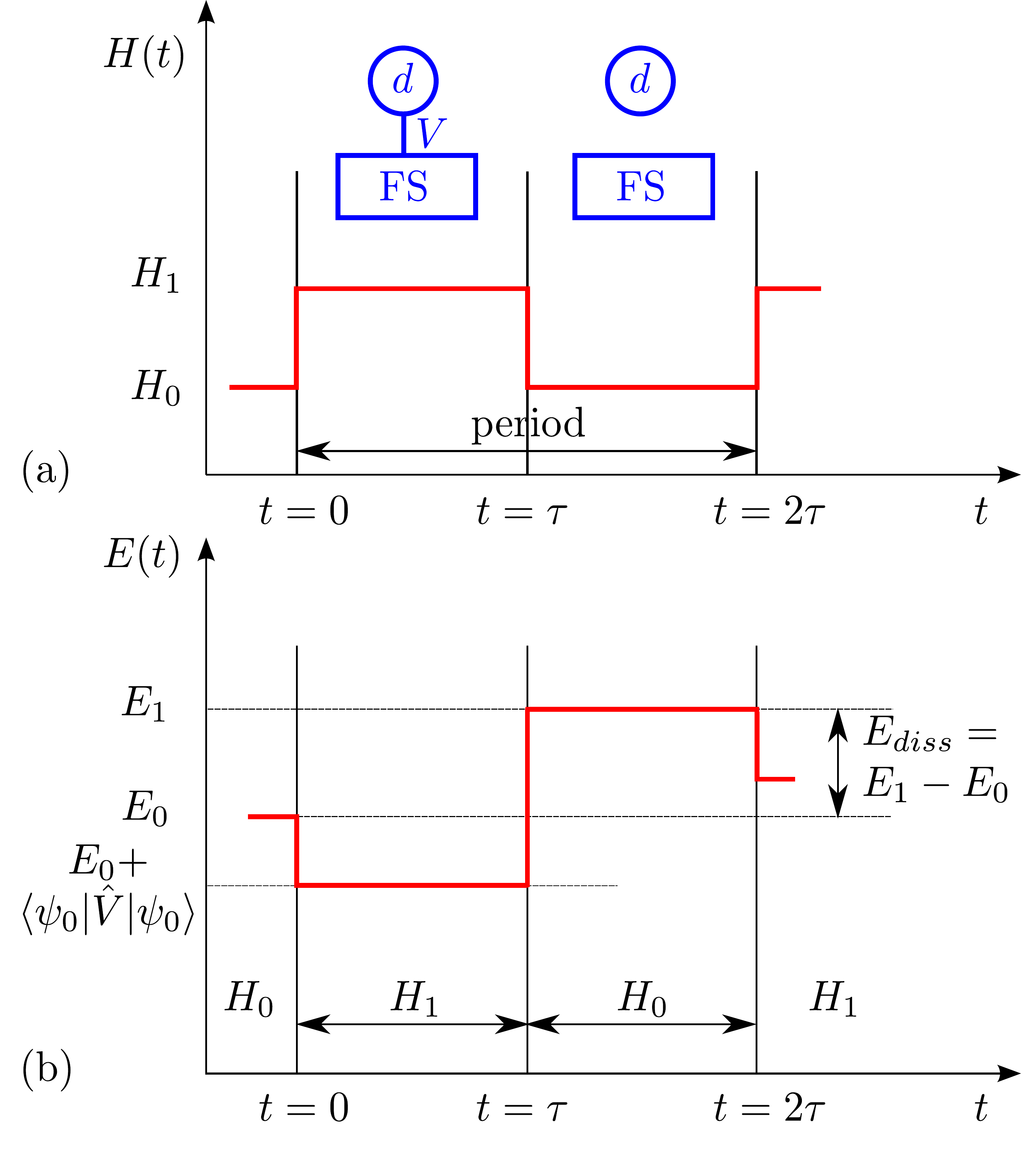}
\caption{(a) Sketch of an ideal system in which the Hamiltonian is mechanically modulated with period $2\tau$.
Starting from $H_0$, there is a switch to Hamiltonian $H_1$, which describes an impurity orbital $d$ coupled to a Fermi sea (FS) 
with density of states $\rho$ by a
hopping
term $V$ leading to a hybridization width $\Gamma=\pi\,V^2\,\rho$
in the first half period. In the second half period the hybridization is switched off, returning to  $H=H_0$.
(b) Evolution of the total energy during a cycle, with indication of the dissipation $E_{diss}$.
}\label{fig_H0H1}
\end{figure}

The difference $\hH_1-\hH_0\equiv \hat{V}$ is a local perturbation.
We shall denote as $\ket{\psi_0}$ and $\mE_0$ the ground state (GS) wavefunction and energy, respectively, of the Hamiltonian $\hH_0$, and as $\ket{\psi_1}$ and $\mE_1$ the corresponding ones of $\hH_1$. \\
A periodic square-wave time variation between $\hH_0$ and $\hH_1$ is assumed, with a very long period 2$\tau$ and a very fast switch time $\tau_{switch}$, much smaller than all 
other time scales.
As sketched in Fig. \ref{fig_H0H1}, a cycle starts at $t=0^-$ with $\hH=\hH_0$, the system in its GS $\ket{\psi_0}$ with energy $E(t<0)=\mE_0$. At $t=0^+$ the Hamiltonian changes into $\hH_1$. 
If the evolution is unitary, the energy becomes
\be
\begin{split}
E(0<t<\tau) \equiv&\, E_0 = \mE_0+\langle \psi_0|\hat V |\psi_0\rangle.
\end{split}
\ee
At $t=\tau^+$ we switch back to
$\hH=\hH_0$ again, so that the energy changes into 
\be
E(\tau<t<2\tau)\equiv E_1 = \bra{\psi(\tau)}\hH_0\ket{\psi(\tau)},\label{def:E_1}
\ee
where 
\[
\ket{\psi(\tau)} = \text{e}^{-i\hH_1 \tau}\;\ket{\psi_0}\,,
\]
and remains constant till the end of the cycle at $t=2\tau$. To proceed on with the successive cycles we define the Hamiltonian $\mathcal{H}_n$ in the time interval $I_n$ 
between $t=n\tau\equiv t_n$ and $t=(n+1)\tau\equiv t_{n+1}$, i.e. 
\be
\mathcal{H}_n = \begin{cases}
\hH_1 & n=\text{even}\,,\\
\hH_0 & n=\text{odd}\,,
\end{cases}
\ee
so that 
\be
\mathcal{H}_{n} - \mathcal{H}_{n-1} = (-1)^n\,\hat{V}\,,\label{Hn.vs.Hn-1}
\ee
and the unitary operator 
\be
\mathcal{U}_n = \exp\Big(-i\,\mathcal{H}_n\,\tau\Big)\,,\label{U_n}
\ee
which evolves the wavefunction from $t_n$ to $t_{n+1}$, namely $\ket{\psi(t_{n+1})} = \mathcal{U}_n\ket{\psi(t_n)}$.
The energy $E_n$ in the same time interval $I_n$ can thus be written as
\be
\begin{split}
E_n =&\, \bra{\psi(t_n)}\mathcal{H}_n\ket{\psi(t_n)}\\ 
=&\, \bra{\psi(t_{n-1})}\mathcal{U}_{n-1}^\dagger\,\mathcal{H}_n\,
\mathcal{U}_{n-1}^\dagga\ket{\psi(t_{n-1})}\\
=&\, E_{n-1} + (-1)^n \bra{\psi(t_n)}\hat{V}\ket{\psi(t_n)}\,,
\end{split}\label{En.vs.En-1}
\ee
assuming $\ket{\psi(t_0)}=\ket{\psi_0}$ and $E_{-1}=\mE_0$. 

\subsection{Dissipated energy}
Let us consider a process that comprises $M$ cycles of duration $2\tau$. 
The dissipated energy, namely the internal energy increase, is therefore 
\be
\begin{split}
\Delta E_\text{diss}(M) &= \bra{\psi(t_{2M-1})}\hH_0\ket{\psi(t_{2M-1})} - 
\bra{\psi_0}\hH_0\ket{\psi_0}\\
&= E_{2M-1}-E_{-1}\,,
\end{split}\label{def:E_diss}
\ee
and is evidently positive. Through Eq.~\eqn{En.vs.En-1} it readily follows that 
\be
\begin{split}
\Delta E_\text{diss}(M) &= E_{2M-1} - E_{-1}\\
&= \sum_{n=0}^{2M-1}\,(-1)^n\,\bra{\psi(t_n)}\hat{V}\ket{\psi(t_n)}\,,
\end{split}\label{def:E_diss-1}
\ee
where, we recall,
\be
\bra{\psi(t_{n+1})}\hat{V}\ket{\psi(t_{n+1})} = \bra{\psi(t_n)}\mathcal{U}_{n}^\dagger\,
\hat{V}\,\mathcal{U}_{n}^\dagga\ket{\psi(t_n)}\,.\label{V-n}
\ee
We can manipulate Eq.~\eqref{def:E_diss-1} to obtain a more manageable expression if $\tau$ is 
much larger than the typical evolution time of the system. 
That is certainly our case, because the action of any tip or other mechanical probe (with KHz to MHz frequency ) is 
many orders of magnitude slower than typical microscopic times.
Then, since $\hat{V}$ is a local operator, we 
can assume thermalisation\cite{thermalization_rmp,thermalization_yukalov}, which implies that the 
expectation value~\eqn{V-n} coincides with the thermal average of $\hat{V}$ over the Boltzmann distribution corresponding to the Hamiltonian $\mathcal{H}_{n}$ at an effective temperature $T_{n}$ 
such that the internal energy is just $E_{n}$.  Since the perturbation $\hat{V}$ that is switched on and off 
is local
in space, involving a volume conventionally assumed to be one,
$E_{n}$ 
lies 
above the ground state energy $\mE_n$ of $\mathcal{H}_{n}$, i.e. $\mE_0$ for odd $n$ and $\mE_1$ otherwise, 
by
a quantity of order 
$\sim n\times O(1)$ as opposed to $\mE_n\sim O(N)$, with $N$ the 
total volume occupied by the system and its Fermi sea, 
so that typically 
$T_n\sim \sqrt{n/N}$. 
It follows that, in the thermodynamic limit and for $M$ finite, $T_n\to 0$ and thus 
Eq.~\eqn{V-n} becomes the expectation value in the ground state of $\mathcal{H}_{n}$ and the 
dissipated energy simplifies into 
\be
\begin{split}
\Delta E_\text{diss}(M) &= M\Big(\bra{\psi_0}\hat{V}\ket{\psi_0} - \bra{\psi_1}\hat{V}\ket{\psi_1}\Big)\\
&\equiv M\, E_\text{diss}\,,
\end{split}
\label{e_diss_gen}
\ee
i.e. $M$ times the energy $E_\text{diss}$ dissipated in a single cycle. 
We could repeat the above arguments at finite temperature $T=\beta^{-1}$ and still find the same expression 
Eq.~\eqn{e_diss_gen} with the GS expectation values replaced by thermal averages, i.e. for $n=0,1$
\be
\begin{split}
\bra{\psi_{n}}\hat{V}\ket{\psi_{n}} &\to
\fract{\text{Tr}\left(\text{e}^{-\beta\,\hH_{n}}\;\hat{V}\right)}
{\text{Tr}\left(\text{e}^{-\beta\,\hH_{n}}\right)}
\equiv \langle \hat{V} \rangle_n\,.
\end{split}
\ee
It is also worth 
proving
that even at finite $T$ the dissipated energy is strictly positive. 
We define a Hamiltonian $\hH(\lambda)=\hH_0+\lambda\,\hat{V}$, with $\lambda\in[0,1]$, and the corresponding free energy and thermal averages, $F(\lambda)$ and $\langle\dots\rangle_\lambda$. 
Through the Hellmann-Feynman theorem the dissipated energy can be written as
\be
\begin{split}
\Delta E_\text{diss}(M,T) &= M\Big(\langle\,\hat{V}\,\rangle_{\lambda=0}
-\langle\,\hat{V}\,\rangle_{\lambda=1}\Big) = M E_\text{diss}(T)\\
&= -M\left(
\fract{\partial F(\lambda)}{\partial \lambda}_{\big|\lambda=1}
-\fract{\partial F(\lambda)}{\partial \lambda}_{\big|\lambda=0}\right)\\
&= -M\int_0^1 d\lambda \;\fract{\partial^2 F(\lambda)}{\partial \lambda^2}\,.
\end{split}\label{e_diss_gen_T}
\ee
Since thermodynamic stability implies that $\partial^2F(\lambda)/\partial\lambda^2 <0$, it follows that 
$E_\text{diss}(T)$ in Eq.~\eqn{e_diss_gen_T} is indeed positive. \\
We stress the importance of the above results: in the limit of an infinite system and of vanishing frequency, i.e. large $\tau$, 
the energy dissipation can be evaluated by an equilibrium calculation, which is evidently more feasible than a full non-equilibrium one.
\begin{figure}[hbt]
\includegraphics[width=0.49\textwidth]{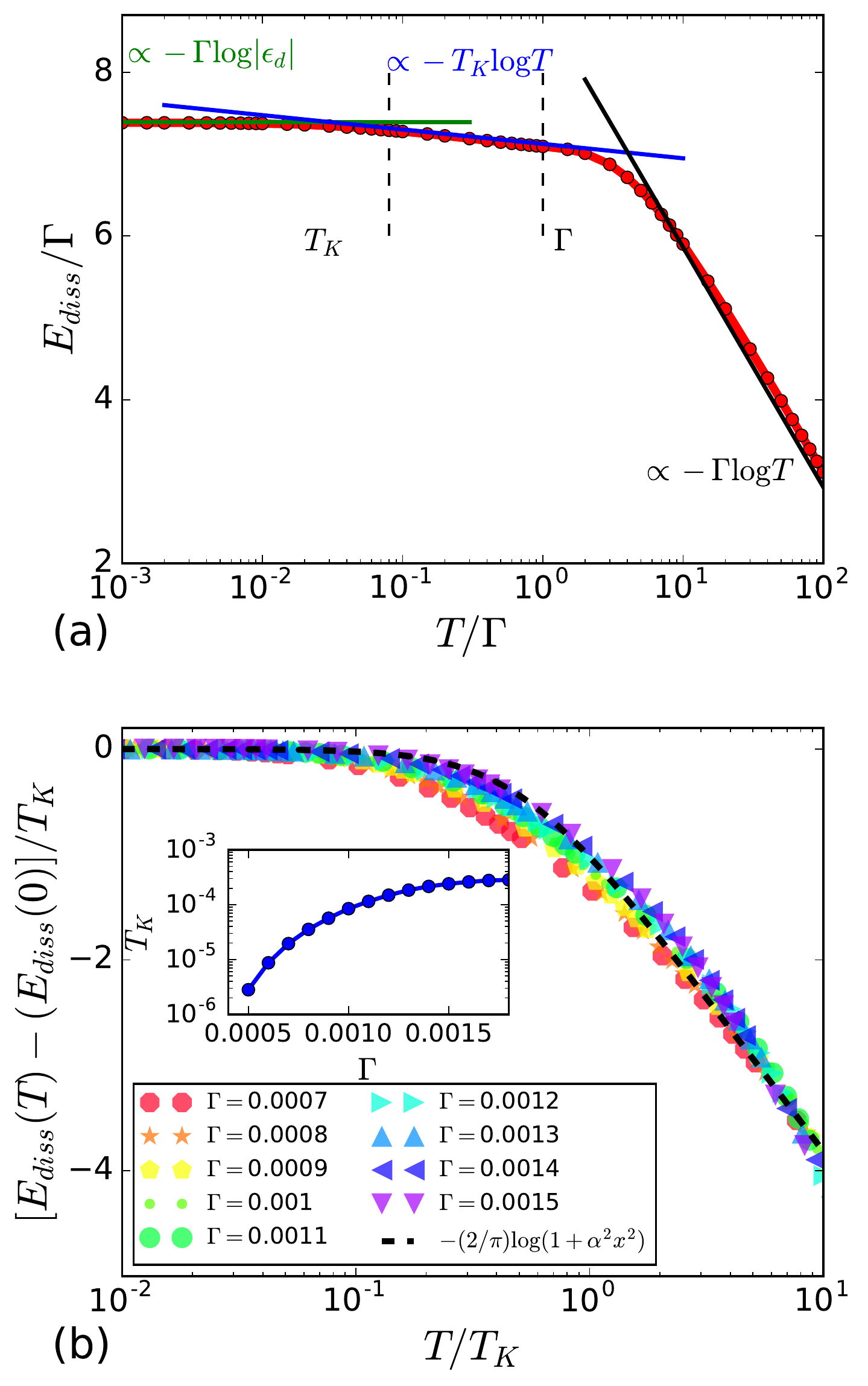}
\caption{(a) Energy dissipation $E_{diss}$ 
per cycle upon switching on and off the hybridization $\Gamma =\pi\,V^2\,\rho\ =0.001$ 
 between a Fermi sea with density of states $\rho$ and an Anderson impurity
with $U=0.01$, $\epsilon_d=-0.005$,
calculated by NRG ($T_K\simeq 8\times 10^{-5}$; when not explicitly indicated, all quantities are in $D$ -- half bandwidth -- units).
Three temperature regimes can be observed: low, $T<T_K$; high, $T>\Gamma$; and intermediate, $T_K<T<\Gamma$. 
(b) In the low temperature regime, $T<T_K$, the dissipation curves for different values of $\Gamma$ can 
be collapsed to a single scaling function Eq.~\eqn{e_diss_tk_t}. In the inset we show $T_K$ extracted from the fit. 
}
\label{fig_nrg}
\end{figure}

\section{Anderson model}\label{sec_nrg}
Since our problem -- calculating the dissipation incurred in a time-dependent but very slow cycle --  can be reduced to calculating 
static local quantities, it can be relatively easily worked out.   
We  model the Kondo 
site
as a single impurity Anderson model (SIAM)\cite{anderson61}
\be
\hH=\hH_\text{bath}+ \hH_d + \hat{V}\,,\label{H_SIAM}
\ee
where $\hH_\text{bath}$ is the Hamiltonian of conduction electrons, which we assume 
to be non-interacting
 with a constant density of states $\rho(\ep)=\rho=1/2D$ for $\ep\in[-D,D]$, and $\rho(\ep)=0$ otherwise. $\hH_d$ is 
the impurity level Hamiltonian,  
\be
\hH_d = \ep_d\,n_d + U\,n_{d\uparrow}\, n_{d\downarrow} - B\, \big(n_{d\uparrow}-n_{d\downarrow}\big)\,,\label{H_d}
\ee
where we 
assume for simplicity
particle-hole symmetry $2\epsilon_d+U=0$. 
We also include the Zeeman splitting 
caused by an external magnetic field $B$ coupled to the $z$ component 
of the impurity spin. The operator $\hat{V}$ is 
an electron hopping term between  
the conduction electrons and the impurity
level, 
giving rise to
the hybridisation energy width 
\be
\Gamma=\pi\,V^2\,\rho\,,\label{Gamma_d}
\ee
and to a Kondo temperature\cite{Hewson}:
\be
T_K \sim \sqrt\frac{\Gamma U}{2} \exp\left(-\frac{\pi U}{8 \Gamma}\right).
\ee
The relative scale of these energies is generally such that  $T_K, B \lesssim \Gamma \ll D,U$. Hereafter $D=1$ is taken as energy unit, and we choose $U=0.01 D$.

\subsection{Switching $\Gamma$ on and off}

We consider the 
periodic switching on and off the hybridisation $\hat{V}$. As explained in introduction, and as portrayed in the cartoon of Fig.1(b), 
this does not precisely reproduce the real effect of a tip on a Kondo impurity, such as sketched in Fig. 1(a);   
but as we shall see it finally does yield  the pertinent information on the Kondo-related dissipation. With the above conventions 
\bea
\hH_0 &=& \hH_\text{bath}+ \hH_d\,,\label{H0-SIAM}\\
\hH_1&=&  \hH_0+\hat V\,,\label{H1-SIAM}
\eea
$\hH_0$ 
describing decoupled bath and impurity, which are instead coupled in $\hH_1$.  
We calculate the Kondo dissipation in the SIAM by cyclically switching on and off the hybridization  through the expression Eq.~\eqn{e_diss_gen_T}, 
valid so long as the system is able to thermalise 
during the
time $\tau$. Strictly speaking,  since the SIAM is exactly solvable by Bethe Ansatz, one 
might question whether thermalisation indeed occurs~\cite{WeymannPRB2015}. However, 
as explained above,
the  time scale $\tau$ set by the period of tip oscillation frequencies is typically lower than 1 MHz, 
a very long time during which electronic thermalization will always occur. 
Another issue is that of electron-electron interactions, which raise the issue of 
how to evaluate Eq.~\eqn{e_diss_gen_T} when 
the bath is made up of interacting electrons. 
Here we shall assume that, if the conduction electron bath can be described within Landau-Fermi liquid theory, 
then we can still evaluate Eq.~\eqn{e_diss_gen_T} through an equilibrium calculation with non-interacting conduction quasiparticles.\\  

Calculations of the expectation values in Eq.~\eqn{e_diss_gen_T} 
are
carried out by the numerical renormalization 
group (NRG) approach~\cite{wilson}, using the ``NRG Ljubljana'' package\cite{nrg_ljubljana} 
with discretisation parameter $\Lambda=2-2.5$,  truncation cutoff $10\,\omega_N$ ($\omega_N=\Lambda^{-N/2}$, $N$ being the $N$-th NRG iteration), $z$-averaging\cite{z_aver} with $z=8$, 
and by means of the full density matrix approach~\cite{weich_dmnrg,costi_fdm}. \\
Fig.~\ref{fig_nrg} reports our results for the energy $E_\text{diss}(T)$ dissipated in a single sub-cycle. 
First we note that the low temperature 
scale of $E_\text{diss}(T)$ is, 
as is to be expected, 
set by the hybridisation $\Gamma$. 
There are three temperature regimes: low, $T<T_K$, intermediate, $T_K<T<\Gamma$,  and high, $T>\Gamma$. 
For $T\ll T_K$, when the Kondo 
screening is fully effective,
its contribution to the dissipation is expected to be a universal scaling function of $T/T_K$. 
We assume that scaling function to be 
assimilated to
that of a resonant level model of width $T_K$, i.e. 
\be\label{e_diss_tk_t}
\fract{E_\text{diss}(T)-E_\text{diss}(0)}{T_K}=-\frac{2}{\pi}\log \left(1+\alpha^2 \,\fract{T^2}{T_K^2}\right).
\ee
A least-square fit with this formula works quite well, see Fig.~\ref{fig_nrg}(b), and provides both an operational definition 
of $T_K$ -- which agrees with the traditional one~\cite{wilson} -- as well as an estimate of the parameter $\alpha \sim 2$. 
Since the thermal average of the hybridisation in the definition of  $E_\text{diss}(T)$, Eq.~\eqn{e_diss_gen_T}, is contributed by degrees of freedom at all energies, 
we must disentangle the Kondo resonance low-energy contribution from the high energy one made mostly by 
larger energy tails of the spectral density, eventually including
the Hubbard side bands. This disentanglement is best operated by the magnetic field 
$B$ in Eq.~\eqn{H_d}, which is known to destroy the Kondo effect\cite{Costi-magnetic-field}. 

\begin{figure}[hbt]
\includegraphics[width=0.4\textwidth]{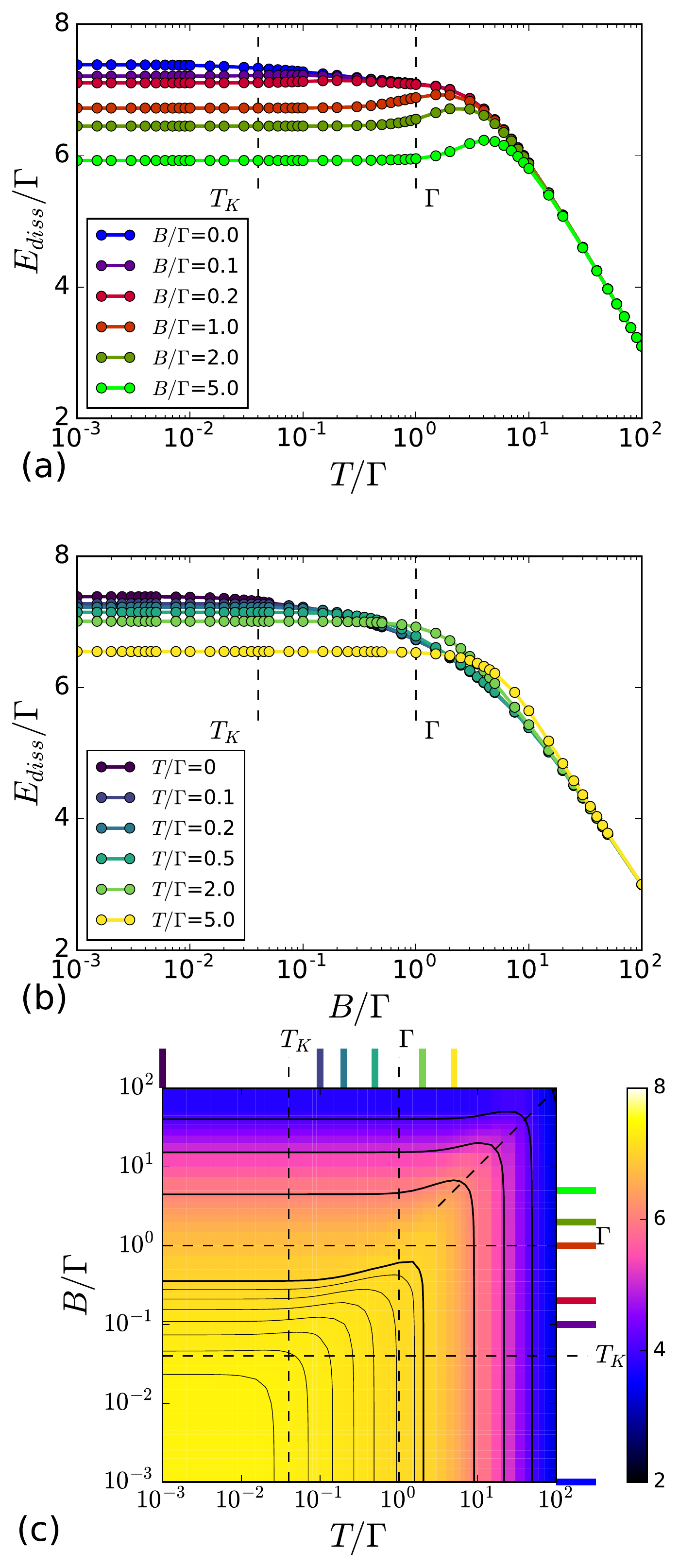}
\caption{(a) Energy dissipation $E_{diss}$ 
per cycle upon switching on and off the hybridization $\Gamma=0.001$ 
 between the same Fermi sea and Anderson impurity as in Fig. \ref{fig_nrg}
in presence of a magnetic field of Zeeman energy $B$:
(a) as a function of $T$ at fixed values of $B$;
(b) as a function of $B$ at fixed values of $T$;
(c) as a function of both $T$ and $B$.
In (c) we show where the cuts at constant $T$ or $B$ have been performed.
}\label{fig_nrg_b}
\end{figure}

\begin{figure}[hbt]
\includegraphics[width=0.4\textwidth]{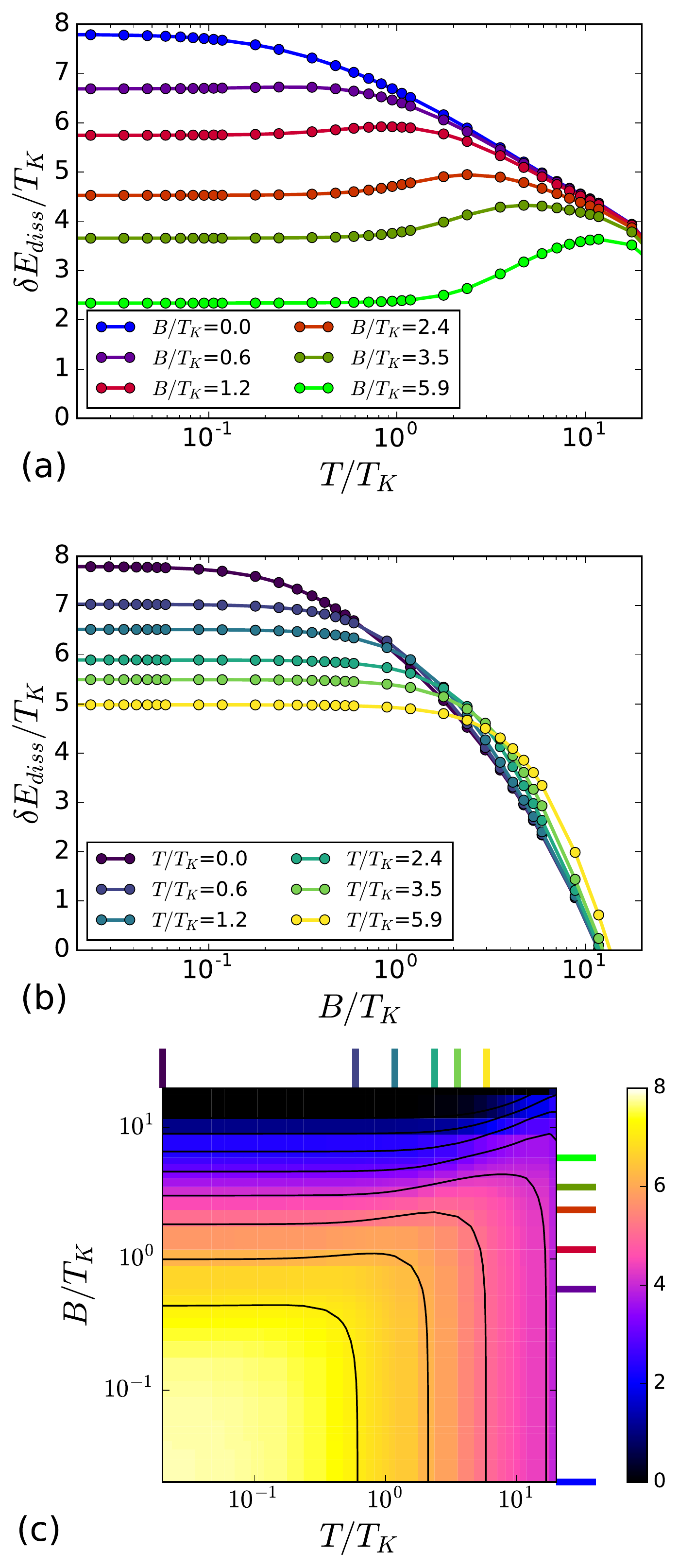}
\caption{
Zoom-in of Fig. \ref{fig_nrg_b}, now showing 
$\delta E_{diss}(T,B)\equiv E_{diss}(T,B)-E_{diss}(0,B=\Gamma)$. The low temperature and low field values give therefore a direct estimate of the dissipation due to Kondo effect.
}\label{fig_nrg_b_bis}
\end{figure}

Confirming that, on increasing the Zeeman splitting we observe 
the disappearance of the $T_K \log T$ behaviour, see Fig.~\ref{fig_nrg_b}, 
replaced by
the growth 
of new peaks at $T \sim  B$, clearly distinguishable 
when $ B\gg \Gamma$. At the meantime we also find a notable reduction of dissipation at low temperature, 
which must therefore entirely come from the magnetic field freezing of the impurity spin and thus from the disappearance of Kondo effect. 
This is more evident in Fig.~\ref{fig_nrg_b_bis}, where we show, at a fixed value of $T_K\simeq 0.1\Gamma$ as  a function of $T$ for different $B$'s, panel (a), and 
as  a function of $B$ for different $T$'s,
panel (b), the difference between 
$E_\text{diss}(T,B)$ and the $T=0$ dissipation $E_\text{diss}(0,\Gamma)$ at field $B=\Gamma\simeq 10T_K$ large enough to kill the Kondo resonance\cite{Costi-magnetic-field}.

\begin{figure}[hbt]
\includegraphics[width=0.45\textwidth]{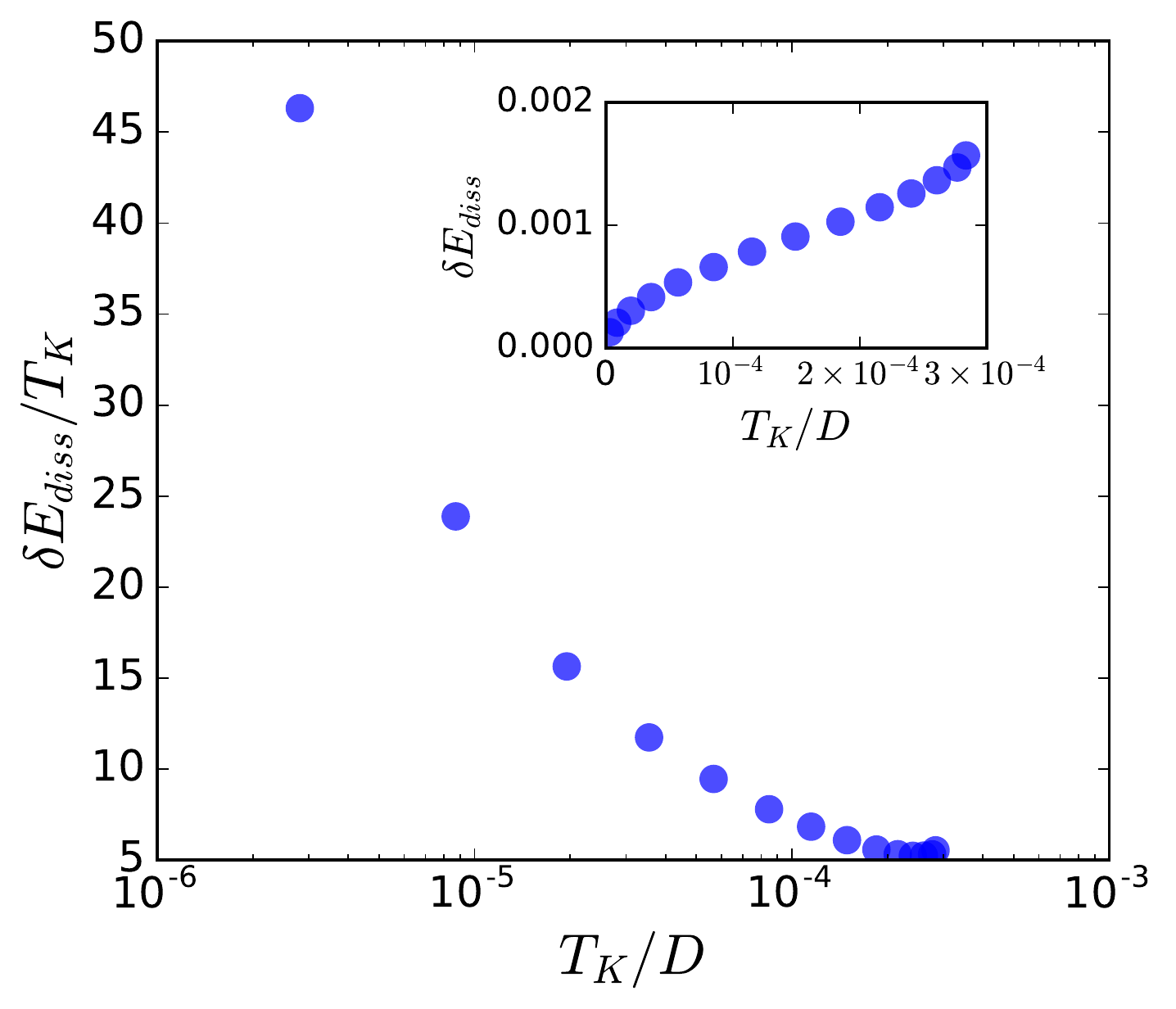}\\
\includegraphics[width=0.45\textwidth]{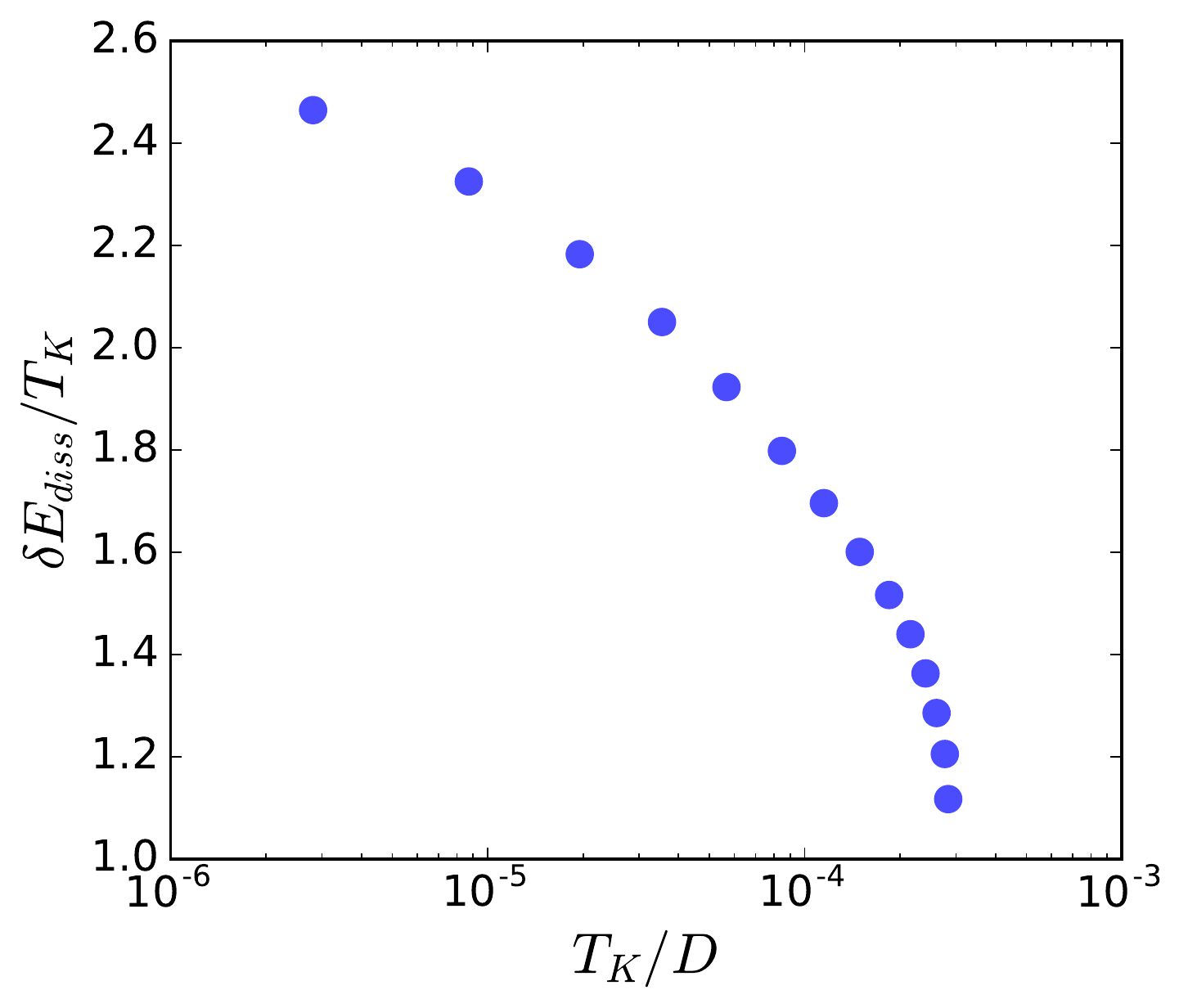}
\caption{Difference $\delta E_\text{diss} = E_\text{diss}(T=0,B=0) 
- E_\text{diss}(T=0,B)$ in units of $T_K$ between the zero temperature values of $E_\text{diss}$ calculated at $B=0$ and at a large $B=\Gamma$, top panel, and at smaller $B=T_K$, bottom panel. 
In the inset we show $\delta E_\text{diss}$ not normalised to $T_K$. 
}\label{fig_nrg_bfit}
\end{figure}

To clarify further the magnitude of the Kondo contribution to dissipation, in Fig.~\ref{fig_nrg_bfit} we 
plot the difference in units of $T_K$ between the zero temperature values of $E_\text{diss}$ calculated at $B=0$ and at $B=\Gamma$ and 
$B=T_K$, top and bottom panel, respectively. The results, both in Fig. \ref{fig_nrg_b_bis} and Fig.~\ref{fig_nrg_bfit} confirm that the Kondo dissipation  
is on the order of $T_K$, as expected, but through a factor 
that grows with decreasing $T_K$ and is typically of order 10 or larger.  

To understand the appearance of this large factor, 
we  observe that the dissipation of  Eq.~\eqn{e_diss_gen_T} may be qualitatively given a spectral representation of the form
\be
E_\text{diss} \sim \int d\ep\,f(\ep)\,\ep\,\rho (\ep)\,,
\ee 
where $f(\ep)$ is the Fermi-Dirac distribution, and $\rho $ is the impurity spectral density. 
While peaked at  $\epsilon =0$, the latter also has long tails away from the Fermi level, which contribute importantly to the integral. 
A large field $B\sim \Gamma$ will filter out the  contributions of all energies $|\ep|\lesssim \Gamma$, 
including both the low-energy, physically meaningful Kondo contribution $\ep\lesssim T_K$  of the Lorentzian Kondo resonance along with the $\big(\log\ep/T_K\big)^{-n}$ slowly decaying tails of the intermediate energy, 
$T_K\lesssim \ep \lesssim \Gamma$, corrections to scaling~\cite{LoganJPCM-2001,z_aver}. The latter,
which contribute to the factor $\delta E_\text{diss}/T_K$ with a strongly singular term $\sim \Gamma/\big(T_K\,\log^2\Gamma/T_K\big)$ as $T_K\to 0$, are actually larger than the former, 
as can be seen by comparing top and bottom panels.\\

Besides these proper Kondo contributions to the dissipation per cycle, 
experiments with an oscillating tip, 
will also cause in general an oscillating shift of the energy of the impurity level $\ep_d$, see Eq.~\eqn{H_d}, from $\ep_d$ to $\ep_d+\delta\ep_d$, taking place at the same time of the on-off switching of Kondo. This 
kind of
modulation may produce an additional dissipation which we might 
designate as
"chemical" in nature. 
When that is important, the 
perturbation $\hat{V}$ 
must be replaced by the more general
\be
\hat{V} \to \hat{V} + \delta\ep_d\,n_d\,.
\ee
However, as long as $\big|\delta\ep_d\big|\ll U$, the effect of a tip-induced level energy modulation simply adds to the 
high-energy contribution to dissipation without altering the low-energy Kondo one. Therefore all the previous results about the way of accessing the Kondo mechanical dissipation by temperature and magnetic field remain valid.

\subsection{Kondo model, switching the exchange $J$}

\begin{figure}[bt]
\includegraphics[width=0.49\textwidth]{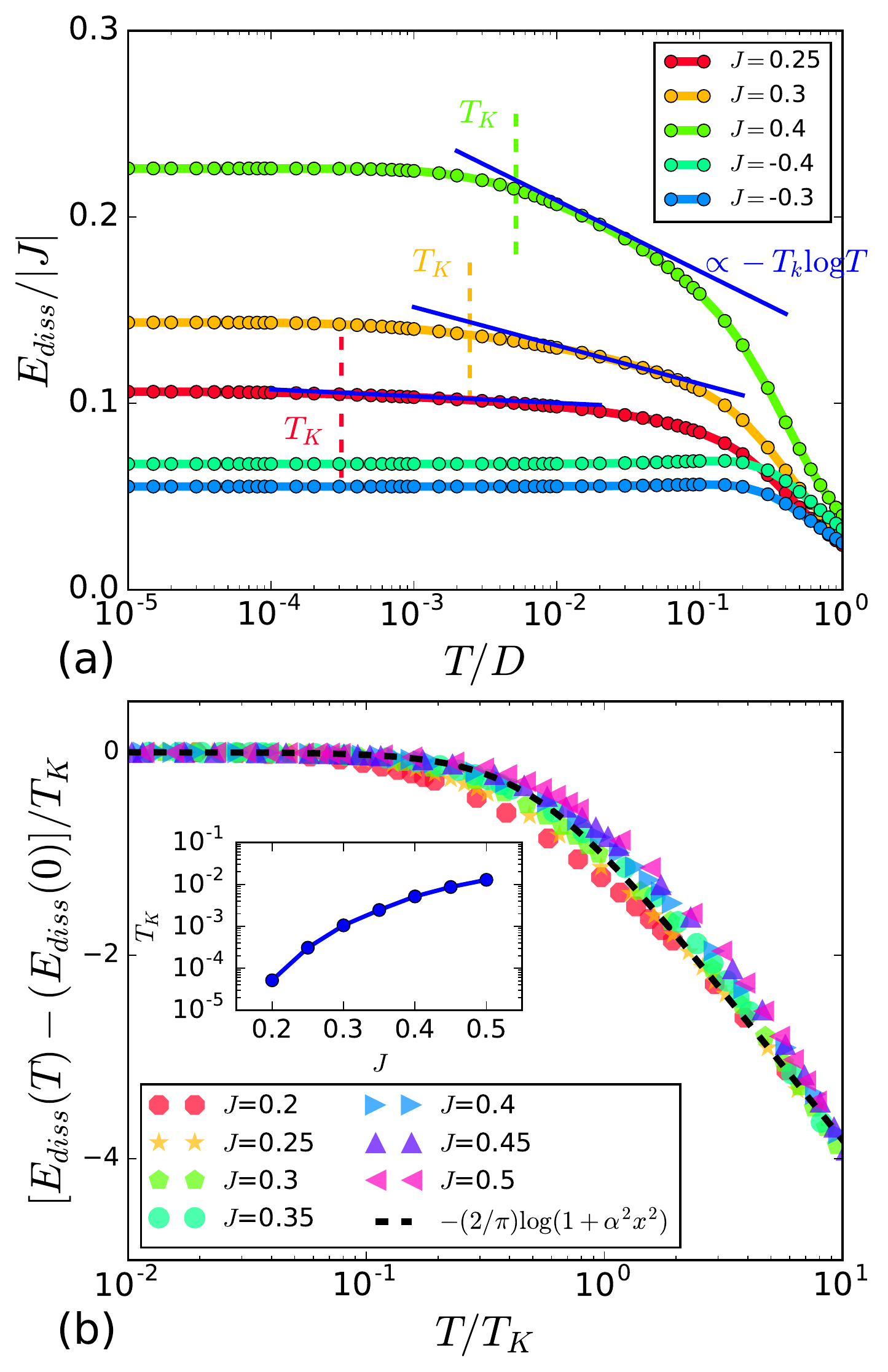}
\caption{(a) Energy dissipation $E_{diss}$ for the single impurity Kondo model  
solved by NRG with different values of $J$, both positive (regular Kondo) and negative (ferromagnetic Kondo). Note the small dissipation and the disappearance of the $T_K \log T$ in the ferro case.
(b) Dissipation around $T_K$ for regular Kondo, $J >0$. As for the Anderson model, it  follows a universal function Eq. \eqref{e_diss_tk_t}, with $\alpha \simeq 2$.
The inset shows $T_K$ as a function of $J>0$. 
}\label{fig_nrg_kondo}
\end{figure}

We have described above the Kondo-related dissipation calculated through an Anderson model. If that result is as general as we surmise, we should be able to recover at least part of it in a simpler Kondo model of the impurity. 
To check that, we apply the same protocol to the case in which the impurity can be effectively regarded as 
a local moment exchange-coupled to the conduction bath, the so-called single impurity Kondo model (SIKM). 
In this model $\hH_0=\hH_\text{bath}$, $\hH_1=\hH_0+\hat V$, where the local perturbation is now
\be
\hat V=J \,\mathbf{S} \cdot \mathbf{s},\label{jss}
\ee
with $\mathbf{S}$ the impurity spin, $\mathbf{s}$ the spin density of the conduction electrons at the impurity site, and $J$ the Kondo exchange which we shall assume to be both positive, antiferromagnetic, and 
negative, ferromagnetic. The energy dissipated in a single sub-cycle $E_\text{diss}$ is still calculated through Eq.~\eqn{e_diss_gen_T} by NRG, as explained above. \\
Fig. \ref{fig_nrg_kondo} present 
the energy dissipation
for this case.
Everything we said for the SIAM holds here as well, except 
for the 
obvious
absence 
of the $\Gamma$ energy scale.
Instead of  $\Gamma$, in this case the total dissipation is now proportional to $J$.
All the same,  in the Kondo regime the decrease of dissipation with temperature is a scaling 
function of $T/T_K$, Eq. \eqref{e_diss_tk_t}, just 
as
for the SIAM. 
The added bonus of 
this simpler SIKM
is that it
gives us the opportunity to investigate the 
dissipation caused by on-off switching of a 
{\it {ferro}}magnetic Kondo effect\cite{Koller2005,Mehta_fmk} by simply setting a negative $J$ in Eq. \eqref{jss}.
%
%
The ferromagnetic Kondo regime
cannot be accessed in the simple SIAM, for which $J>0$, unless one 
resorts to
a more complicated microscopic modelling \cite{io_fmk}. 
In ferromagnetic Kondo,  $J<0$, 
the total dissipation is still on the order of $|J|$, 
but,
as expected, no Kondo energy scale 
appears
anymore.
The dissipation is constant up to $T\simeq |J|$, where it has a slight increase (absent in the antiferromagnetic case),
and finally decays at higher temperatures like in the antiferromagnetic case.

\section{Discussion and Conclusions} \label{sec_conclusions}
We introduce 
and discuss
the concept of dissipation  connected with switching on and off an impurity Kondo effect, such as could be realized by oscillating STM or AFM tips, but not only.  
We then present a direct scheme to estimate and calculate it. 
In the many-body Anderson model, a model where our desired quantities can to our knowledge be calculated only numerically,
we show that square-wave-like switching on and off the impurity hybridization, 
although different from the physical action of a tip, permits the extraction of the specific Kondo-related dissipation.
The dissipation per cycle connected with creation  and destruction of the Kondo cloud is identified by the 
change of dissipation caused by
temperature and magnetic field.\\ 

Results show that the Kondo part of dissipation per cycle is, not surprisingly,  proportional to the Kondo temperature $T_K$.
However, the proportionality factor, of order one for $T$ and $B$ of order $T_K$, may soar 
to a   surprisingly larger factor, of order 5-10, for larger $T$ or larger $B$, reflecting the role of spectral density tails even far from $E_F$.
Even though the mechanisms intervening in a real experimental setup will likely be more complicated than assumed here, 
our treatment can be a good starting point for the description of dissipation of surface-adsorbed impurities perturbed by AFM or STM tips. 
In particular, our assumption of a simple hybridization quench is a crude one: in a vibrating tip experiment, other
parameters of the Hamiltonian will change too, and the model itself should be expanded to account for additional degrees of freedom, 
for example the coupling between mechanical and electronic ones.

With these experiments in mind we should mention how the predicted Kondo dissipation per cycle compares with the sensitivity of these systems. 
The most recent "pendulum AFM"~\cite{Kisiel2011} realized with tips of Q-factor $4.8 10^5$, vibrating with frequency near 5 KHz,  
force constant $k \sim 0.03$N/m and amplitude $A\sim 5$nm can reach the extreme sensitivity $\pi k A^2 \sim 10^{-5}$eV/cycle. 
Our calculated Kondo dissipation, Figs. \ref{fig_nrg_b_bis} and \ref{fig_nrg_bfit},  is generally of order $k_BT_K$ per cycle at low temperature.
Typically of order $10^{-3}$ eV or bigger, this is  
substantially
larger than the sensitivity, making the predicted Kondo dissipation easily accessible and measurable. 
Even if stronger tip-impurity contact interactions than typical pendulum AFM ones may be needed in order to cause the Kondo switch, 
and despite the fact that chemical or mechanical dissipation sources must obviously be subtracted, 
this estimate indicates that Kondo dissipation should be quite relevant to the real world. 
It is therefore hoped that our results will encourage experiments in the new area at the border between nanofriction and many body physics.

\section{Acknowledgements}
Sponsored by ERC MODPHYSFRICT Advanced Grant No. 320796 and ERC FIRSTORM Advanced Grant No. 692670. 
We thank Rok \v Zitko for help with the NRG code and for useful discussions.

\bibliography{biblio.bib}

\end{document}